\documentstyle[12pt]{article}
\newcommand{\beq}{\begin{equation}}
\newcommand{\eeq}{\end{equation}}
\newcommand{\bea}{\begin{eqnarray}}
\newcommand{\eea}{\end{eqnarray}}

\begin{document}
\setcounter{page}{0}
\topmargin 0pt
\oddsidemargin 5mm
\renewcommand{\thefootnote}{\fnsymbol{footnote}}
\newpage
\setcounter{page}{0}
\begin{titlepage}
\begin{flushright}
QMW 93-16
\end{flushright}
\begin{flushright}
hep-th/9305106
\end{flushright}
\vspace{0.5cm}
\begin{center}
{\large {\bf Some Comments On Lie-Poisson Structure Of Conformal Non-Abelian
Thirring Models}} \\
\vspace{1.8cm}
\vspace{0.5cm}
{\large Oleg A. Soloviev
\footnote{e-mail: soloviev@V1.PH.QMW.ac.uk}\footnote{Work supported by
S.E.R.C.}}\\
\vspace{0.5cm}
{\em Physics Department, Queen Mary and Westfield College, \\
Mile End Road, London E1 4NS, United Kingdom}\\
\vspace{0.5cm}
\renewcommand{\thefootnote}{\arabic{footnote}}
\setcounter{footnote}{0}
\begin{abstract}
{The interconnection between self-duality, conformal
invariance and Lie-Poisson structure of the two dimensional non-abelian
Thirring model is investigated in the framework of the hamiltonian method.}
\end{abstract}
\vspace{0.5cm}
\centerline{May 1993}
 \end{center}
\end{titlepage}
\newpage
\section{Introduction}
 In the present paper we will discuss the interconnection between the Siegel
symmetry [1], self-duality, conformal invariance and Lie-Poisson structure of
the two dimensional non-abelian Thirring model [2]. Our motivations in
considering
the Thirring model come from string theory. It has been observed that a
given conformal Thirring model should correspond to a certain compactification
in string theory [3,4]. Therefore, the space of all conformal Thirring models
seems to be a good candidate to describe the space of all symmetric string
vacua, which could form the space of conformal backgrounds
appropriate to the formulation of background independent string field theory
[5].

The remarkable universality of Thirring models originates from the invariance
of these theories under the symmetry first introduced by W. Siegel [1]. The
Siegel symmetry is a crucial property of self-dual fields in two dimensions
[1]. We will show that self-duality provides some clues in understanding the
geometrical quantization of the non-abelian conformal Thirring models.

\section{The equivalence between non-abelian fermionic and bosonic Thirring
models}

One of the manifestations of universality in the Thirring model is the
equivalence between its fermionic and bosonic formulations [2].
The fermionic Thirring model action is given by
\begin{equation}
S_F=\int d^2x(\bar\psi_L\partial\psi_L +\bar\psi_R\bar\partial\psi_R+
S_{a\bar a}J^a_LJ^{\bar a}_R),\end{equation}
where $\psi_L$ and $\psi_R$ are Weyl spinors (in general carrying a flavor)
transforming as
the fundamental representations of given groups $G_L$ and $G_R$ respectively.
The last term
in (1)
describes the general interaction between fermionic currents $J^a_L
=\bar\psi_Lt^a\psi_L,\;J^{\bar a}_R=\bar\psi_Rt^{\bar a}\psi_R$, where
$t^a,\;t^{\bar
a}$ are the generators in the Lie algebras ${\cal G}_L,\;{\cal G}_R$. $S_{a\bar
a}$ is a
coupling constant matrix.

The action of the bosonic Thirring model is formulated as follows
\begin{equation}
S_B=\int [L_L(k_L,g_L)+L_R(k_R,g_R)+L_{int}(g_L,g_R;S)],\end{equation}
where these three terms respectively are given by
\begin{eqnarray}
4\pi
L_L(k_L,g_L)&=&-k_L[(1/2)tr_L|g^{-1}_Ldg_L|^2+(i/3)d^{-1}tr_L(g^{-1}_Ldg_L)^3],
\nonumber\\
4\pi
L_R(k_R,g_R)&=&-k_R[(1/2)tr_R|g_R^{-1}dg_R|^2+(i/3)d^{-1}tr_R(g^{-1}_Rdg_R)^3],\\
L_{int}(g_L,g_R;S)&=&-(k_Lk_R/4\pi)tr_Ltr_Rg^{-1}_L\partial g_L\cdot S\cdot
dg_Rg^{-1}_R,\nonumber\end{eqnarray}
with the coupling $S$ belonging to the direct product ${\cal G}_L\otimes{\cal
G}_R$. Here the fields $g_L$ and  $g_R$ take their values in the Lie groups
$G_L$
and $G_R$, respectively, $k_L,\;k_R$ are central elements in the affine
algebras $\hat{\cal G}_L,\;\hat{\cal G}_R$. The symbols $tr_L,\;tr_R$ indicate
tracing over the group indices of $G_L,\;G_R$.

Classically the theories (1) and (2) are inequivalent, whatever
conditions we may impose upon them. However, at the quantum level the fermionic
and
bosonic non-abelian Thirring models become indistinguishable under the
following conditions: 1) the two Weyl spinors $\psi^i_R$ and $\psi_L^{\bar i}$
carry the flavor indices $i=1,...,k_R$ and $\bar i=1,...,k_L$; 2) the
coupling constant matrix $S$ in eqs. (1), (2) and (3) is reversible; 3)
the fields $g_L$ and $g_R$ are left and right moving scalars
respectively [2]. When these conditions are fulfilled the statistical sums of
the two
models are identical [2]
\begin{equation}
{Z_B(k_L,k_R;S/4\pi)\over Z_B(k_L,k_R;0)}={Z_F(k_L,k_R;S)\over Z_F(k_L,k_R;0)},
\end{equation}
where $Z_B,\;Z_F$ are defined via usual partition functions of the given two
dimensional models.

Apparently, in the limit $S=0$ the identity (4) contains no useful
information. It is not surprising because as we
demonstrated in [6] in order to fermionize the WZNW models (or $S=0$ Thirring
model) with arbitrary
levels, we have to use the fermionic Thirring model at the so-called isoscalar
Dashen-Frishman conformal points, not at $S_{ab}=0$.
Meanwhile, when $S\neq 0$, the identity (4) is very fruitful since allows us
to establish an equivalence between the conformal points of the fermionic and
bosonic versions of the Thirring model as well as to clarify its geometrical
meaning [7].

\section{Conformal points of the Thirring model}

There are considerable merits of Thirring models which make them especially
interesting in finding the appropriate unification of both conformal field
theories and massive integrable models. Therefore, it would be illuminating
if one could explore the Thirring model at all the possible values of the
couplings $S_{ab}$. However, this seems to be beyond our present analytical
abilities. Most of the difficulty resides in the highly non-linear
character
of the current-current interaction of the Thirring theory. Given our present
knowledge, the theory is tractable only when it possesses either affine
symmetry or quantum group symmetry (which might turn out to be a sort of
deformation of the former.) In this paper we will not discuss the quantum
group symmetry of Thirring models but rather affine symmetries. We will show
that affine symmetries are intimately related to the conformal invariance of
the
Thirring model. The non-abelian Thirring model has been shown to have at
least two types of
conformal points for different values of the Thirring coupling constants.

The conformal points belonging to the first type are called Higgs conformal
points [8]. They may appear in the theory only when $k_L\neq k_R$. For example,
in the simplest case $G_L=G_R=G$ and $S=\sigma\;t^a\otimes t^a$ conformal
invariance holds at the following values of $\sigma$ [8]
\begin{equation}
\sigma_n^{L,R}=\left({k_Lk_R\over (k_L+c_2(G))(k_R+c_2(G))}\right)^{-n}\sigma^
{L,R}_0,
\end{equation}
where $\sigma^{L,R}_0=1/k_{L,R}$; $c_2(G)$ is a quadratic Casimir operator
eigenvalue referring to the adjoint representation of the group $G$;
$n=0,1,2,...,\infty.$ Interestingly, all Higgs conformal points  share
the same Virasoro central charges [8]
\begin{eqnarray}
c(k_L,k_R,\sigma_n^L)=c(k_L,k_R,\sigma^L_0)=
\left({k_L\over k_L+c_2(G)/2}+{k_R-k_L\over
k_R-k_L+c_2(G)/2}\right)\dim G,\nonumber \\ & & \\
c(k_L,k_R,\sigma_n^R)=
c(k_L,k_R,\sigma^R_0)=\left({k_R\over k_R+c_2(G)/2}+{k_L-k_R\over
k_L-k_R+c_2(G)/2}\right)\dim G.\nonumber\end{eqnarray}
Meanwhile, the second derivative of the $c$-function of Zamolodchikov [9] at
the conformal points (5) varies from one conformal point to another and goes
to zero in the limit $n\rightarrow \infty$ [8].

{}From the equivalence between the fermionic and bosonic statistical sums, the
Higgs conformal points of the bosonic Thirring model should be the critical
points in the fermionic theory. Note that it could be seen also by
straightforward calculations of the fermionic partition functions at the given
values of the coupling constants. However in the bosonic case the analysis is
much simpler
[8].

In its turn the fermionic formulation of the Thirring model turns out to be
easier in finding the so-called Dashen-Frishman conformal points [6] which
are the natural generalization of the isoscalar conformal points discovered by
Dashen and Frishman two decades ago [10].

At the Dashen-Frishman conformal points the non-linear Thirring model can be
quantized non-perturbatively in the framework of operator quantization [6].
The procedure amounts to quantizing the classical equation of motion
\begin{equation}
\partial\psi_L=-2S_{a\bar a}J^{\bar a}_Rt^a\psi_L.\end{equation}
We have shown in [6] that the r.h.s of eq. (7) can be well defined at the
quantum level if
the coupling matrix $S_{ab}$ satisfies the so-called Virasoro master equation
[11]. This result being purely non-perturbative seems somewhat mysterious,
because
given the conformal points, we cannot employ even feeble perturbative
arguments to
justify the conformal invariance of the Thirring model at the quantum level.
Therefore some more sophisticated arguments are required. We are going to
demonstrate that the Dashen-Frishman conformal points are
quite natural for the bosonic Thirring model despite the analysis
being a little more cumbersome, compared to the fermionic version.

\section{Dashen-Frishman conformal points and Lie-Poisson structure of the
Thirring model}

To clarify the nature of the Dashen-Frishman conformal points we have to
consider more carefully the structure of the classical bosonic non-abelian
Thirring
model. The action of the bosonic Thirring model can be presented in the
geometrical form [7]
\begin{equation}
S_B=\int \alpha_L\;+\;\int \alpha_R\;+\; S_{int}, \end{equation}
where $\alpha_L$ and $\alpha_R$ are canonical one-forms associated to the
canonical symplectic
structures on the orbits of the affine groups $\hat G_L$ and $\hat G_R$
respectively [12,13].
\begin{equation}
d\alpha_L=\omega_L,\;\;\;\;\;\;\;\;d\alpha_R=\omega_R.\end{equation}
The symplectic forms $\omega_L$ and $\omega_R$ define canonical
variables and their Poisson brackets. We will show that the last term in eq.
(8) is
a hamiltonian in the phase space with the symplectic forms given above.

To this end, we rewrite
the actions for the free WZNW models in the first order form [15]
\begin{eqnarray}
A_L=-(1/4\gamma_L)\int
tr_L[\partial_0g_Lg_L^{-1}J_{0L}-(1/2)(J^2_{0L}+J^2_{1L})]dxdt+ WZ_L,\;\;\;
\gamma_L=\pi/k_L,\nonumber\\& &\\
A_R=-(1/4\gamma_R)\int
tr_R[\partial_0g_Rg_R^{-1}J_{0R}-(1/2)(J^2_{0R}+J^2_{1R})]dxdt+WZ_R,\;\;\;
\gamma_R=\pi/k_R,\nonumber\end{eqnarray}
where $WZ_{L,R}$ are WZ-terms, i.e. the second terms in the r.h.s. of
$L_{L,R}$ in (3). The WZ-terms are linear in $\partial_0g_{L,R}$ and
therefore can be considered as functionals of $g_L,\;g_R$ alone [15].
We have also used the notations
\begin{eqnarray}
J_{1L}=\partial_xg_L\cdot g_L^{-1},\nonumber\\& &\\
J_{1R}=\partial_xg_R\cdot g_R^{-1}.\nonumber\end{eqnarray}

The variation of the canonical 1-forms in $A_L,\;A_R$ leads us to the
symplectic forms
\begin{eqnarray}
\Omega_L=(1/4\pi)\int tr_L(dg_Lg_L^{-1}\wedge
dJ_{0L}+J_{1L}dg_Lg_L^{-1}\wedge dg_Lg_L^{-1})dxdt,\nonumber\\& &\\
\Omega_R=(1/4\pi)\int tr_R(dg_Rg_R^{-1}\wedge
dJ_{0R}+J_{1R}dg_Rg_R^{-1}\wedge dg_Rg_R^{-1})dxdt.\nonumber\end{eqnarray}
These symplectic forms are a little different from those in (9). The difference
is the same as between usual coordinates and light-cone coordinates. We can
find the Poisson brackets for variables $g_L,\;J_{0L},\;g_R,\;J_{0R}$ by
inverting $\Omega_L,\;\Omega_R$. We
find
\begin{eqnarray}
\{g^1_L(x),g^2_L(y)\}&=&0,\nonumber\\
\{J^1_{0L}(x),g^2_L(y)\}&=&-2\gamma_LC_Lg_L^2(y)\delta(x-y),\\
\{J^1_{0L}(x),J^2_{0L}(y)\}&=&-\gamma_L[J^1_{0L}(x)-J^1_{1L}(x)-J^2_{0L}(x)+
J^2_{1L}(x),C_L]\delta(x-y),\nonumber\end{eqnarray}
and similar ones for $g_R,\;J_{0R}$. Here $\{A^1(x),B^2(y)\}$ denotes
the $2\dim G_L\times 2\dim G_L$ matrix of all Poisson brackets of $\dim
G_L\times\dim G_L$ matrices $A$ and $B$, arranged in the same fashion, as in
the product of matrices
\begin{eqnarray}
A^1=A\otimes I \nonumber\end{eqnarray}
and
\begin{eqnarray}
B^2=I\otimes B;\nonumber\end{eqnarray}
$C_L$ is a constant $2\dim G_L\times 2\dim G_L$ matrix given by
\begin{eqnarray}
C_L={\sum_a}\;t^a\otimes t^a.\nonumber\end{eqnarray}

Note that the actions $A_L,\;A_R$ are equivalent to the actions of the WZNW
models upon use of the equations of motion for $J_{0L},\;J_{0R}$. Therefore,
the sets
$(g_L,\;J_{0L})$ and $(g_R,\;J_{0R})$ describe the independent
canonical variables in $A_L,\;A_R$. For our aims, however, it is more
convenient to introduce
new coordinates which are
\begin{equation}
g_L,\;\;\;\;L=(1/2)g_L^{-1}(J_{0L}+J_{1L})g_L\end{equation}
for the $A_L$-theory, and
\begin{equation}
g_R,\;\;\;\;R=(1/2)(J_{0R}-J_{1R})\end{equation}
for the $A_R$-theory. The Poisson brackets for the new variables follow from
eqs.
(12) and can be also obtained by inverting the symplectic forms in (9). In
particular [15]
\begin{eqnarray}
\{L^1(x),L^2(y)\}&=&
(\gamma_L/2)[C_L,L^1(x)-L^2(y)]\delta(x-y)+\gamma_LC_L\delta'(x-y),\nonumber\\&
&\\
\{R^1(x),R^2(y)\}&=&(\gamma_R/2)[C_R,R^1(x)-R^2(y)]\delta(x-y)+\gamma_RC_R
\delta'(x-y).\nonumber\end{eqnarray}

Now the interaction term in eq. (8) can be seen as a hamiltonian in the
phase space of variables (14), (15)
\begin{equation}
S_{int}=\int dx^+dx^-\;{\cal H},\end{equation}
with the Hamiltonian density
\begin{equation}
{\cal H}=-{4\pi\over \gamma_L\gamma_R}\langle S,L\otimes R\rangle,
\end{equation}
where $S={\sum_{a\bar a}}S_{a\bar a}t^a\otimes t^{\bar a}$.
The hamiltonian $H^+=\int dx^-{\cal H}$
 describes the evolution of the system in the
$x^+$-direction. The hamiltonian equation for $g_R$ is given by
\begin{equation}
\partial_+g_R+(4\pi/\gamma_L)(tr_LS\cdot L)g_R=0.\end{equation}
Thus by solving this equation, we can express $L$ in terms of
$g_R$. Due to the symmetry between $g_L$ and $g_R$, the equation for $g_L$
has to be as follows
\begin{equation}
\partial_-g_L+(4\pi/\gamma_R)g_L(tr_R S\cdot R)=0.\end{equation}
In fact the last equation could be derived as a hamiltonian equation with the
hamiltonian
\begin{equation}
H^-=\int dx^+{\cal H}.
\end{equation}
Hence, eqs. (19), (20) give rise to the explicit
expressions for $L$ and $R$ in terms of $g_R$ and $g_L$. Thus, the action
\begin{equation}
A=A_L+A_R+\int dx^+dx^- {\cal H}\end{equation}
becomes a functional of the fields $g_L$ and $g_R$ alone.
Now it is not very difficult to see that the given functional coincides with
the action of the bosonic Thirring model  upon use of the Siegel constraints.
A noteworthy fact is that the Siegel constraints similar to
(19) and (20) appear
in the Thirring model as the self-duality conditions [1, 14] via
introduction of
lagrangian multipliers [1], whereas in the geometric formulation these same
constraints appear as hamiltonian equations of motion without any auxiliary
fields.

Now we may try to quantize the Thirring model by the hamiltonian method. The
method will work as long as the algebraic Poisson structure will be preserved.
In
other words, to gain the advantage of geometric quantization, we
have to promote the affine algebra Poisson brackets (16) to the
quantum level. Certainly it will be the case, if the quantum
fields $g_L,\;g_R$ are elements of the representations of the affine algebras
$R,\;L$ respectively. As a byproduct
the theory should be conformally invariant, since the Virasoro
algebra belongs to the enveloping algebra of the affine algebra. It means that
the hamiltonian quantization of the Thirring model will be consistent as long
as conformal symmetry will be present.

Therefore, for consistent quantization, we have to impose
the following quantum equations of motion
\begin{eqnarray}
\left[L_{-1},g_R\right]=-(4\pi /\gamma_L)tr_L:S\cdot L\cdot g_R:,\nonumber\\ &
&\\
\left[\bar L_{-1},g_L\right]=-(4\pi/\gamma_R)tr_R:g_L\cdot S\cdot R:,\nonumber
\end{eqnarray}
Here the double dots :: denote normal ordering between the currents $L,\;R$ and
their affine primary fields $g_R,\;g_L$ respectively. The brackets [,] are
understood as quantum analogues of the classical Poisson brackets. The
operators
$L_{-1}$ and $\bar L_{-1}$ are to be the generators of translations. By
definition
\begin{equation}
L_{-1}=\oint {dz\over 2\pi i}T(z),\;\;\;\;\;\bar L_{-1}=\oint{d\bar z\over 2\pi
i}\bar T(\bar z),\end{equation}
with $T(z)$ and $\bar T(z)$ the holomorphic and antiholomorphic components of
the energy-momentum tensor in the conformal Thirring model.

Let us suppose that $\hat{\cal G}_L=\hat{\cal G}_R$. Then we can construct the
following operators
\begin{eqnarray}
T(z)=L_{ab}\tilde L^a\tilde L^b,\;\;\;\;\tilde L=(1/\gamma_L)L,\nonumber\\& &\\
\bar T(\bar z)=L_{ab}\tilde R^a\tilde R^b,\;\;\;\;\tilde
R=(1/\gamma_R)R,\nonumber\end{eqnarray}
which one can use to obtain equations (23). It is not hard to check that
eqs. (23) are fulfilled with the given $T$ and $\bar T$ provided
$L_{ab}=4\pi S_{ab}$. Moreover, to be
components of the conformal energy-momentum tensor, the given
operators must form two copies of the Virasoro algebra. In the full analogy
with the fermionic Thirring model [6], we can prove that the operators $T$ and
$\bar T$ form the Virasoro algebras if and only if the matrix $L_{ab}$
satisfies the Virasoro master equation [11]. Thus, we have a one-to-one
correspondence between the Dashen-Frishman conformal points of the fermionic
Thirring model and the self-consistence conditions of the bosonic Thirring
model.

\section{Conclusion}

In summary, the bosonic Thirring model at the Dashen-Frishman conformal points
possesses the explicit affine symmetries realized by the Poisson
algebras (16). These symmetries are inherited by the quantum theory from the
Lie-Poisson algebra of the classical geometrical formulation. At the same
time, due to the constraints (19), (20), we
could expect certain algebraic structures for the operators
$\partial g_Rg_R^{-1}$ and
$g_L^{-1}\bar\partial g_L$ despite them not playing any role in the
quantization. We might guess that these operators should be generators
of quantum algebras with the $r$-matrices depending on $S_{ab}$. It is
very tempting to suppose that among all the solutions of the Virasoro master
equation there should be solutions to the Yang-Baxter equation. Then at the
given
conformal points we might find as a byproduct the realization of the quantum
algebras in
terms of Thirring models.

It is also amusing that at the Dashen-Frishman conformal points, the
interaction term in
the lagrangian becomes a truly marginal operator. Therefore it would be very
interesting to investigate the flows between the underlying free WZNW models
and the Thirring models
at the Dashen-Frishman conformal points. Then we probably could understand the
nature of the finite
conformal deformations and $c>1$ string models. In this respect, the
flows between Dashen-Frishman conformal points and Higgs conformal points are
also very important.

The hope is that the realization of this program might shed some more light on
the space of two dimensional hamiltonian theories [16].

\par \noindent
{\em Acknowledgement}: I would like to thank J. Gates, M. Green, C. Hull,
E. Ramos, K. G. Selivanov and A. Zubarev  for fruitful discussions. I also
thank S. Thomas for the careful reading of the manuscript

\end{document}